\documentstyle[12pt]{article}

\begin{document}

\author{M.\ Apostol  \\ 
Department of Theoretical Physics,\\Institute of Atomic Physics,
Magurele-Bucharest MG-6,\\POBox MG-35, Romania\\e-mail: apoma@theor1.ifa.ro}
\title{On the low-dimensional solids and their melting }
\date{}
\maketitle

\begin{abstract}
It is shown that the one- and two-dimensional solids may exist and melt.
\end{abstract}

Suppose that we have $N$ identical atoms, of mass $M$ each, arranged along a
rectilinear chain; suppose that the atoms interact through a
nearest-neighbours potential, and we are interested in small atomic
displacements $u_i\;,\;i=1,2,...N,$ of immaterial polarization, around
equidistant equilibrium positions $x_i=i\cdot a,$ where $a$ is the lattice
constant. Within the harmonic approximation with the elastic force constant $%
K$ this motion is described by the hamiltonian 
\begin{equation}
\label{one}H=\sum_i\frac 1{2M}p_i^2+\frac 12K\sum_{\langle ij\rangle
}(u_i-u_j)^2\;\;\;, 
\end{equation}
where $p_i$ is the conjugate momentum, $\left[ p_i,u_j\right] =-i\hbar
\delta _{ij}\;.$ With cyclic boundary conditions we have 
\begin{equation}
\label{two}u_i=\frac 1{\sqrt{N}}\sum_ku_ke^{ikx_i}\;\;\;, 
\end{equation}
$k=(2\pi /Na)\cdot ${\it integer}, $-\pi /a<k<\pi /a,\;u_k^{\dagger
}=u_{-k}\;,$%
\begin{equation}
\label{three}p_i=M\stackrel{\cdot }{u_i}=\frac 1{\sqrt{N}%
}\sum_kp_ke^{-ikx_i}\;\;\;, 
\end{equation}
$p_k=M\stackrel{\cdot }{u}_{-k}\;,\;\left[ p_k,u_{k^{^{\prime }}}\right]
=-i\hbar \delta _{kk^{^{\prime }}}\;$, and the hamiltonian $\left( 1\right) $
becomes 
\begin{equation}
\label{four}H=\sum_k\left( \frac 1{2M}p_k^{\dagger }p_k+\frac 12M\omega
_k^2u_k^{\dagger }u_k\right) \;\;\;, 
\end{equation}
where 
\begin{equation}
\label{five}\omega _k^2=\frac{4K}M\left( 1-\cos ka\right) \;\;; 
\end{equation}
it is brought into the diagonal form 
\begin{equation}
\label{six}H=\sum_k\hbar \omega _k\left( n_k+1/2\right) \;\;\;, 
\end{equation}
where $n_k=a_k^{\dagger }a_k\;,$ by introducing the phonon operators $a_k\;,$%
\begin{equation}
\label{seven}u_k=\sqrt{\frac \hbar {2M\omega _k}}\left( a_k+a_{-k}^{\dagger
}\right) \;\;\;,\;\;\;p_k=i\sqrt{\frac{\hbar M\omega _k}2}\left(
a_k^{\dagger }-a_{-k}\right) \;\;. 
\end{equation}
The mean square deviation of the atomic displacements in the phonon vacuum 
\begin{equation}
\label{eight}\overline{u_i^2}=\frac 1N\sum_k\overline{u_k^{\dagger }u_k}%
=\frac 1N\sum_k\frac \hbar {2M\omega _k}\left( 2\overline{n}_k+1\right)
=\frac a{2\pi }\int_0^{\pi /a}dk\cdot \frac \hbar {M\omega _k} 
\end{equation}
diverges logarithmically at $k=0\;.$ This infrared divergence, in various
contexts, is currently invoked to rule out the existence of the
one-dimensional solid.\cite{Peierls}$^{,}$\cite{Landau}

\medskip\ 

However, the mean square deviation of the lattice constant 
\begin{equation}
\label{nine}
\begin{array}{c}
\overline{\left[ x_i-x_j-\overline{\left( x_i-x_j\right) }\right] ^2}=%
\overline{\left( u_i-u_j\right) ^2}=\frac 1N\sum_k\overline{u_k^{\dagger }u_k%
}\cdot 2\left( 1-\cos ka\right) = \\ =\frac 1N\sum_k\frac \hbar {M\omega
_k}\left( 1-\cos ka\right) \left( 2\overline{n}_k+1\right) =\frac a\pi
\int_0^{\pi /a}dk\cdot \frac \hbar {M\omega _k}\left( 1-\cos ka\right) 
\end{array}
\end{equation}
is finite, which indicates that the solid exists, only its center of mass
fluctuates immeasurably. Obviously, the thermodynamics is meaningless in
this case.

\medskip\ 

The question one should ask is whether the phonons do exist in the
one-dimensional solid subjected to the actual experimental conditions. These
always imply an uncertainty $u$ in the localization of the center of mass of
the rigid lattice, and the constraint under which we should look for phonons
reads 
\begin{equation}
\label{ten}\sum_i\overline{u_i^2}=Nu^2\;\;. 
\end{equation}
The magnitude of $u$ can not be smaller than $a/2$\ , unless the crystal is
destroyed, but usually it is much larger than the lattice constant $a.$

\medskip\ 

Introducing the Lagrange multiplier $\omega _0^2$ and setting 
\begin{equation}
\label{eleven}{\cal H}=H-\frac 12M\omega _0^2\sum_i\left( u^2-u_i^2\right) 
\end{equation}
we get straightworwardly 
\begin{equation}
\label{twelve}{\cal H}=\sum_k\left( \frac 1{2M}p_k^{\dagger }p_k+\frac
12M\Omega _k^2u_k^{\dagger }u_k\right) -\frac 12M\omega _0^2Nu^2\;\;\;, 
\end{equation}
where 
\begin{equation}
\label{thirteen}\Omega _k^2=\omega _0^2+\omega _k^2\;\;\;, 
\end{equation}
{\it i.e.} the phonon spectrum acquires a threshold frequency $\omega _0$
which acts as an infrared cut-off. In addition, the ground-state energy 
\begin{equation}
\label{fourteen}E_0=\overline{{\cal H}}=\sum_k\frac 12\hbar \Omega _k-\frac
12M\omega _0^2Nu^2 
\end{equation}
has a minimum with respect to $\omega _0^2,$%
\begin{equation}
\label{fifteen}\frac{\partial E_0}{\partial \omega _0^2}=\sum_k\frac \hbar
{4\Omega _k}-\frac 12MNu^2\;\;\;, 
\end{equation}
which is exactly the constraint $\left( 10\right) $ ; indeed, 
\begin{equation}
\label{sixteen}\sum_i\overline{u_i^2}=\sum_k\overline{u_k^{\dagger }u_k}%
=\sum_k\frac \hbar {2M\Omega _k}=Nu^2 
\end{equation}
coincides with $\left( 15\right) $. Assuming a Debye spectrum $\omega _k=ck$
, where $c=a(2K/M)^{\frac 12}$ is the sound velocity, and introducing the
Debye wavevector $k_D=\pi /a$ and the Debye frequency $\omega _D=ck_D$ , we
get from $\left( 16\right) $%
\begin{equation}
\label{seventeen}\int_0^{\omega _D/\omega _0}dx\cdot \frac 1{\sqrt{1+x^2}}=%
\frac{2M\omega _D}\hbar u^2\;\;. 
\end{equation}
The right-hand side of $\left( 17\right) $ is, typically, very large, so
that we get the threshold frequency 
\begin{equation}
\label{eighteen}\omega _0\cong 2\omega _D\cdot e^{-\frac{2M\omega _D}\hbar
\cdot u^2} 
\end{equation}
which is extremely small.

\medskip\ 

The phonon partition function for ${\cal H}$ given by $\left( 12\right) $
reads 
\begin{equation}
\label{nineteen}Z=\prod_k\frac{e^{-\hbar \Omega _k/2T}}{1-e^{-\hbar \Omega
_k/T}}=e^{-(F+\frac 12M\omega _0^2Nu^2)/T}\;\;\;, 
\end{equation}
where $F$ is the free energy; its minimum $\partial F/\partial \omega
_0^2=0\;$gives 
\begin{equation}
\label{twenty}\sum_k\frac \hbar {2M\Omega _k}\left( 2\overline{n}_k+1\right)
=Nu^2\;\;\;, 
\end{equation}
{\it i.e.} again the constraint $\left( 10\right) ,$ at finite temperatures.
Equation $\left( 20\right) $ reads 
\begin{equation}
\label{twentyone}\int_0^{\omega _D/\omega _0}dx\cdot \frac 1{\sqrt{1+x^2}%
}\cdot \coth \left( \frac{\hbar \omega _0}{2T}\sqrt{1+x^2}\right) =\frac{%
2M\omega _D}\hbar \cdot u^2\;\;\;, 
\end{equation}
whose solution for $T\rightarrow 0$ is $\left( 18\right) $ and 
\begin{equation}
\label{twentytwo}\omega _0\cong \frac \pi {2M\omega _Du^2}\cdot T 
\end{equation}
for $T\rightarrow \infty \;;\;$this is again an extremely small frequency
for any normal temperature, and $\left( 22\right) $ may be used
satisfactorily for any finite temperature. The constraint $\left( 10\right) $
introduces therefore an extremely small infrared cut-off $\omega _0$ in the
phonon spectrum, according to $\left( 13\right) $, slightly depending on
temperature, whose effect is practically unobservable. We remark that the
mean square deviation of the distance between any pair of atoms 
\begin{equation}
\label{twentythree}\overline{\left( u_i-u_j\right) ^2}=\frac 1N\sum_k\frac
\hbar {2M\Omega _k}\cdot 2\left\{ 1-\cos \left[ ka\left( i-j\right) \right]
\right\} \cdot \left( 2\overline{n}_k+1\right) 
\end{equation}
is always smaller than $4u^2$ , according to $\left( 20\right) $ , so that
the (on-diagonal) crystalline long-range order does exist in one dimension.

\medskip\ 

In this connection the question of melting of the one-dimensional solid is
meaningfull. The elastic force constant $K$ in $\left( 1\right) $ is a
function of $a$ ; actually it is a function of the nearest-neighbours
distance $a+u_i-u_j$ ; more than this, it is a periodic function with the
period $a$ , whose integral extended to the whole crystal vanishes.
Consequently, it may be expanded as a Fourier series of the reciprocal
vectors of the lattice, and we may write it as 
\begin{equation}
\label{twentyfour}K(a+u_i-u_j)=K\cos \left[ G\left( u_i-u_j\right) \right]
+...\;\;\;, 
\end{equation}
where $G=2\pi /a$ . Assuming vanishing fluctuations we may approximate the
cosine in $\left( 24\right) $ by its average 
\begin{equation}
\label{twentyfive}\overline{\cos \left[ G\left( u_i-u_j\right) \right] }%
=e^{-\frac 12\delta ^2}\;\;\;, 
\end{equation}
where $\delta ^2=\overline{\left[ G\left( u_i-u_j\right) \right] ^2}$ , and
keep the first $G$-term in $\left( 24\right) $. The sound velocity changes,
therefore, from $c$ to $c\cdot \exp \left( -\delta ^2/4\right) $ , and we
get a mean-field theory known as the self-consistent harmonic
approximation.\ Remark that it is $\overline{\left( u_i-u_j\right) ^2}$
which enters the theory, not $\overline{u_i^2}$ .\cite{Fukuyama} On the
other hand 
\begin{equation}
\label{twentysix}\delta ^2=G^2\overline{\left( u_i-u_j\right) ^2}=\frac{G^2}%
N\sum_k\frac \hbar {M\Omega _k}\left( 1-\cos ka\right) \cdot \left( 2%
\overline{n}_k+1\right) \;\;\;, 
\end{equation}
and we may neglect $\omega _0$ here and approximate $1-\cos ka$ by $\left(
ka\right) ^2/2$ over the whole integration range. With the Debye spectrum $%
\left( 26\right) $ becomes 
\begin{equation}
\label{twentyseven}\delta ^2=\frac{2\pi ^2G^2}{\hbar M\omega _D^3}\cdot
T^2\cdot e^{\frac 34\delta ^2}\cdot \int_0^{\frac{\hbar \omega _D}{2T}%
e^{-\delta ^2/4}}dx\cdot x\coth x\;\;. 
\end{equation}
Threre is a critical temperature beyond which this equation has no solution
anymore; it is given by the tangent point of the curves described by the two
sides of $\left( 27\right) $ , which may be approximated by 
\begin{equation}
\label{twentyeight}1\cong \frac{2\pi ^2G^2}{\hbar M\omega _D^3}\cdot
T^2\cdot \frac 34e^{\frac 34\delta ^2}\cdot \int_0^{\frac{\hbar \omega _D}{2T%
}e^{-\delta ^2/4}}dx\cdot x\coth x\;\;\;, 
\end{equation}
whence, together with $\left( 27\right) $, yields 
\begin{equation}
\label{twentynine}\delta ^2\cong \frac 43\;\;. 
\end{equation}
The high-temperature limit of $\left( 27\right) $ reads now 
\begin{equation}
\label{thirty}\delta ^2=\frac{\pi ^2G^2}{M\omega _D^2}\cdot T\cdot e^{\frac
12\delta ^2}\;\;\;, 
\end{equation}
whence we get the melting temperature (not to be mistaken for the freezing
temperature\cite{Ramakrishnan}) 
\begin{equation}
\label{thirtyone}T_m=\frac{4M\omega _D^2}{3\pi ^2G^2e^{2/3}}=\frac 1{3\pi
^2e^{2/3}}\cdot Mc^2\cong 0.017\cdot Mc^2\;\;. 
\end{equation}
It is interesting to compare $\overline{\left( u_i-u_j\right) ^2}=a^2/3\pi
^2 $ at the melting point given by $\left( 29\right) $ with the square of
the lattice constant $a^2$\ . The melting temperature given by $\left(
31\right) $ indicates a sharp transition where the crystal gets soft and can
no longer bear phonons. The investigation of the validity of this mean-field
theory may indicate, in general, a continuous transition, with a variable
range of crystallinity, both in magnitude and orientation.\cite{LandauLif}
Indeed, the well-known\cite{LandauLif} argument for the existence of the
multi-phased one-dimensional thermodynamical systems can be applied here for
the breaking up of the inter-atomic bonds between neighbouring atoms. For a
one-dimensional crystalline solid consisting of $N$ atoms, with an energy $%
\psi $ for each bond, the equilibrium is reached for a certain, finite value
of the concentration $n/N\ll 1$ given by 
\begin{equation}
\label{thirtytwo}n=N\cdot e^{-\psi /T}\;\;\;, 
\end{equation}
where $n$ is the total number of broken bonds. This shows that the
one-dimensional crystalline solid is unstable, at any finite temperature,
with respect of the breaking up of the interatomic bonds (including the
bonds with the substrate that ensure the constraint given by $\left(
10\right) $). This instability may be viewed as a continuous transition
toward a state with a variable range of crystallinity. We should remark,
however, that the concentration given by $\left( 32\right) $ is extremely
small for any realistic values of the bonding energy $\psi $ and normal
temperatures.

\medskip\ 

The constraint $\left( 10\right) $ applies to all the atoms in the lattice,
as, for example, to a lattice with a basis in the unit cell. However, the
optical phonons do not contribute practically to the translational
displacement of the lattice, so that the threshold frequency $\omega _0$
(appearing now in the whole phonon spectrum) is practically unchanged. The
constraint $\left( 10\right) $ applies also to solids of any dimensionality,
and we shall discuss now, for the sake of reference, the case of a cubic
three-dimensional Bravais lattice. The hamiltonian of the atomic vibrations
can be written as 
\begin{equation}
\label{thirtythree}H=\sum_{\alpha i}\frac 1{2M}p_{\alpha i}^2+\frac
12\sum_{\alpha \beta \langle ij\rangle }G_{\alpha \beta }\left( {\bf r}_i-%
{\bf r}_j\right) u_{\alpha i}u_{\beta j}\;\;\;, 
\end{equation}
where $\alpha ,\beta $ are the cartesian labels of the components of the
displacement vector ${\bf u}_i\;,\;{\bf r}_i$ are the equilibrium positions,
and $G_{\alpha \beta }$ is the matrix of the elastic force constants.\ A
Fourier expansion brings $H$ into 
\begin{equation}
\label{thirtyfour}H=\sum_{\alpha {\bf k}}\frac 1{2M}\cdot p_{\alpha {\bf k}%
}^{\dagger }p_{\alpha {\bf k}}+\frac 12\sum_{\alpha \beta {\bf k}}G_{\alpha
\beta }\left( {\bf k}\right) \cdot u_{\alpha {\bf k}}^{\dagger }u_{\beta 
{\bf k}}\;\;\;,, 
\end{equation}
where 
\begin{equation}
\label{thirtyfive}G_{\alpha \beta }\left( {\bf k}\right) =\sum_jG_{\alpha
\beta }\left( {\bf r}_i-{\bf r}_j\right) \cdot e^{i{\bf k}\left( {\bf r}_i-%
{\bf r}_j\right) }\;\;\;, 
\end{equation}
$j^{^{\prime }}$s being the nearest neighbours of $i$ . A canonical
transform diagonalizes $G_{\alpha \beta }\left( {\bf k}\right) $ into $%
G_\alpha \left( {\bf k}\right) $ , where we use the same label $\alpha $ for
polarizations. In addition, we assume a Debye, isotropic, identical spectrum
for all the polarizations, 
\begin{equation}
\label{thirtysix}G_\alpha \left( {\bf k}\right) =M\omega _k^2=Mc^2k^2\;\;\;. 
\end{equation}
We remark that the constraint $\left( 10\right) $, which now reads 
\begin{equation}
\label{thirtyseven}\sum_i\overline{{\bf u}_i^2}=Nu^2 
\end{equation}
is left unchanged under the canonical transform. Under these conditions we
get the spectrum $\Omega _k^2=\omega _0^2+\omega _k^2$ and $\left( 37\right) 
$ gives 
\begin{equation}
\label{thirtyeight}\sum_{\alpha {\bf k}}\frac \hbar {2M\Omega _k}\cdot \coth
\left( \frac{\hbar \Omega _k}{2T}\right) =Nu^2\;\;. 
\end{equation}
Making use of the Debye wavevector $k_D=\left( 6\pi ^2\right) ^{1/3}/a$ ,
and the Debye frequency $\omega _D=ck_D\;,\;\left( 38\right) $ becomes 
\begin{equation}
\label{thirtynine}\frac{3\hbar \omega _0^2}{2M\omega _D^3}\cdot
\int_0^{\omega _D/\omega _0}dx\cdot \frac{x^2}{\sqrt{1+x^2}}\cdot \coth
\left( \frac{\hbar \omega _0}{2T}\sqrt{1+x^2}\right) =\frac{u^2}3\;\;\;. 
\end{equation}
For $T\rightarrow 0$ the only solution of $\left( 39\right) $ is $\omega
_0=0 $ and the zero-point fluctuations $u_0^2=9\hbar /4M\omega _D\;\left(
\ll a^2\right) \;;\;$for\ $T\rightarrow \infty $ we find $\omega _0=0$
again, and $u^2=9T/M\omega _D^2$ , which varies extremely slowly with the
temperature, and may be used for any normal range of temperatures.
Therefore, there is no frequency cut-off on the phonon spectrum in three
dimensions, and the lattice develops its own fluctuations.

\medskip\ 

In order to compute the melting temperature we use again the self-consistent
harmonic approximation,according to which 
\begin{equation}
\label{forty}
\begin{array}{c}
G_{\alpha \beta }\left( 
{\bf r}_i-{\bf r}_j\right) \Longrightarrow G_{\alpha \beta }\left( {\bf r}_i-%
{\bf r}_j+{\bf u}_i-{\bf u}_j\right) = \\ \sum_{{\bf G}}G_{\alpha \beta
}\left( {\bf G}\right) \cos \left[ {\bf G}\left( {\bf u}_i-{\bf u}_j\right)
\right] \Longrightarrow \sum_{{\bf G}}G_{\alpha \beta }\left( {\bf G}\right)
e^{-\frac 12\overline{\left[ {\bf G}\left( {\bf u}_i-{\bf u}_j\right)
\right] ^2}}\;\;. 
\end{array}
\end{equation}
On the other hand 
\begin{equation}
\label{fortyone}
\begin{array}{c}
{\bf G}\left( {\bf u}_i-{\bf u}_j\right) =\sum_\alpha G_\alpha \left(
u_{\alpha i}-u_{\alpha j}\right) = \\ =\frac 1{\sqrt{N}}\sum_{\alpha {\bf k}%
}G_\alpha u_{\alpha {\bf k}}\left[ 1-e^{i{\bf k}\left( {\bf r}_j-{\bf r}%
_i\right) }\right] e^{i{\bf kr}_i} 
\end{array}
\end{equation}
and 
\begin{equation}
\label{fortytwo}\overline{\left[ {\bf G}\left( {\bf u}_i-{\bf u}_j\right)
\right] ^2}=\frac 1N\sum_{\alpha {\bf k}}G_\alpha ^2\cdot \overline{%
u_{\alpha {\bf k}}^{\dagger }u_{\alpha {\bf k}}}\cdot 2\left\{ 1-\cos \left[ 
{\bf k}\left( {\bf r}_i-{\bf r}_j\right) \right] \right\} \;\;. 
\end{equation}
For a cubic lattice the sum in $\left( 42\right) $ does not depend on $%
\langle ij\rangle $ and it may be approximated by 
\begin{equation}
\label{fortythree}\overline{\left[ {\bf G}\left( {\bf u}_i-{\bf u}_j\right)
\right] ^2}\cong \frac{G^2a^2}N\sum_{{\bf k}}\frac \hbar {2M\omega _k}\cdot
k_x^2\cdot \left( 2\overline{n}_k+1\right) \;\;\;, 
\end{equation}
or 
\begin{equation}
\label{fortyfour}\overline{\left[ {\bf G}\left( {\bf u}_i-{\bf u}_j\right)
\right] ^2}=G^2\frac{8\left( 6\pi ^2\right) ^{2/3}}{\hbar ^3M\omega _D^5}%
\cdot T^4\int_0^{\hbar \omega _D/2T}dx\cdot x^3\coth x\;\;\;. 
\end{equation}
We may therefore keep the first ${\bf G}^{^{\prime }}$s in $\left( 40\right) 
$ and, denoting the corresponding $\left( 44\right) $ by $\delta ^2$ , we
have 
\begin{equation}
\label{fortyfive}G_{\alpha \beta }\left( {\bf r}_i-{\bf r}_j\right)
\Longrightarrow G_{\alpha \beta }\left( {\bf r}_i-{\bf r}_j\right) \cdot
e^{-\frac 12\delta ^2}\;\;\;, 
\end{equation}
\begin{equation}
\label{fortysix}c\Longrightarrow c\cdot e^{-\frac 14\delta ^2}\;\;\;, 
\end{equation}
and a similar relation for $\omega _D$ . Equation $\left( 44\right) $
becomes 
\begin{equation}
\label{fortyseven}\delta ^2=G^2\frac{8\left( 6\pi ^2\right) ^{2/3}}{\hbar
^3M\omega _D^5}\cdot T^4\cdot e^{\frac 54\delta ^2}\cdot \int_0^{\frac{\hbar
\omega _D}{2T}e^{-\frac 14\delta ^2}}dx\cdot x^3\coth x\;\;\;, 
\end{equation}
whose critical solutions are given by 
\begin{equation}
\label{fortyeight}\delta ^2\cong 4/5\;\;. 
\end{equation}
The finite-temperature estimation of $\left( 47\right) $ is 
\begin{equation}
\label{fortynine}\delta ^2=\frac{G^2a^2}{3Mc^2}\cdot T\cdot e^{\frac
12\delta ^2}\;\;\;, 
\end{equation}
whence the melting temperature 
\begin{equation}
\label{fifty}T_m=\frac 3{5\pi ^2e^{2/3}}\cdot Mc^2\cong 0.04\cdot Mc^2\;\;; 
\end{equation}
this is the Lindemann's law of melting.\cite{Lindemann} At the melting
temperature the fluctuations in the atomic positions are 
\begin{equation}
\label{fiftyone}u_m^2=\frac 9{M\omega _D^2}\cdot T_m\cong 0.025a^2\;\;\;, 
\end{equation}
according to $\left( 39\right) $ , and we may see that their ratio to the
zero-point fluctuations is given by 
\begin{equation}
\label{fiftytwo}u_m^2/u_0^2=4T_m/\hbar \omega _D\cong 0.04\cdot \frac{Mca}%
\hbar \;\;. 
\end{equation}
\medskip\ 

We shall discuss now briefly the two-dimensional case. Under the same
assumptions as those employed for the three-dimensional solid the constraint
equation $\left( 39\right) $ becomes 
\begin{equation}
\label{fiftythree}\frac{\hbar \omega _0}{M\omega _D^2}\int_0^{\omega
_D/\omega _0}dx\cdot \frac x{\sqrt{1+x^2}}\cdot \coth \left( \frac{\hbar
\omega _0}{2T}\sqrt{1+x^2}\right) =\frac{u^2}2\;\;\;, 
\end{equation}
where $\omega _D=ck_D,\;k_D=2\sqrt{\pi }/a\;.$ For $T\rightarrow 0$ this
equation has the solution $\omega _0=0$ and the zero-point fluctuations $%
u_0^2=2\hbar /M\omega _D.$ For $T\rightarrow \infty $ we get 
\begin{equation}
\label{fiftyfour}\omega _0\cong \omega _D\cdot e^{-\frac{M\omega _D^2}{4T}%
\cdot u^2}\;\;\;, 
\end{equation}
{\it i.e. }the two-dimensional solid may exist at finite temperatures,\cite
{Mermin} under the constraint $\left( 10\right) $. The cut-off frequency
given by $\left( 54\right) $ is extremely small and varies very slowly with
the temperature; it may be used for any finite temperature range (remark
that it vanishes for $T\rightarrow 0$ ), and its effects are practically
unnoticeable. The melting may be treated similarly with the
three-dimensional case. Instead of $\left( 47\right) $ we get now 
\begin{equation}
\label{fiftyfive}\delta ^2=G^2\frac{16\pi }{\hbar ^2M\omega _D^4}\cdot
T^3\cdot e^{\delta ^2}\int_0^{\frac{\hbar \omega _D}{2T}e^{-\frac 14\delta
^2}}dx\cdot x^2\cdot \coth x\;\;\;, 
\end{equation}
whose critical solution is given by 
\begin{equation}
\label{fiftysix}\delta ^2\cong 1\;\;. 
\end{equation}
The finite-temperature integral in $\left( 55\right) $ gives 
\begin{equation}
\label{fiftyseven}\delta ^2=\frac{2\pi ^2}{Mc^2}\cdot T\cdot e^{\frac
12\delta ^2}\;\;\;, 
\end{equation}
whence the melting temperature 
\begin{equation}
\label{fiftyeight}T_m=\frac 1{2\pi ^2e^{1/2}}\cdot Mc^2\cong 0.03\cdot
Mc^2\;\;. 
\end{equation}
We stress again the possibility of a continuous transition,\cite{Kosterlitz}
including various structural phases.

\medskip\ 

Similar considerations as those presented here apply for highly anisotropic
solids. A natural question arises of what happens for quasi-low dimensional
solids, as for a slab or a rod of thickness $d$ . The qualitative picture
given here remains unchanged, but, of course, the quantitative results are
altered. The main feature is the occurrence of a cross-over temperature
toward a three-dimensional behaviour, of the order of $\hbar c/d$ ; for $%
d=N_ta$ this temperature is about $10^2/N_t\;$Kelvins, for a typical sound
velocity $c\sim 10^3\;m/s\;.$


\begin{thebibliography}{9}
\bibitem{Peierls}  R.\ Peierls, Helv.\ Phys. Acta {\bf 7 }, Suppl.2 , 81
(1934).

\bibitem{Landau}  L. Landau, ZhETF {\bf 7} 627 (1937) (Phys. Z. Sowjet. {\bf %
11 }545 (1937)).

\bibitem{Fukuyama}  H.\ Fukuyama and P.\ M.\ Platzmann, Solid State Commun. 
{\bf 15 }677 (1974).

\bibitem{Ramakrishnan}  T.\ V.\ Ramakrishnan and M.\ Yussouff, Phys.\ Rev. B%
{\bf 19} 2775 (1979).

\bibitem{LandauLif}  L.\ Landau and E.\ Lifshitz, Physique Statistique, Mir,
Moscow (1967), last paragraph.

\bibitem{Lindemann}  A.\ Lindemann, Z.\ Phys. {\bf 11} 609 (1910).

\bibitem{Mermin}  N.\ D.\ Mermin, Phys.\ Rev. {\bf 176} 250 (1968).

\bibitem{Kosterlitz}  I.\ M.\ Kosterlitz and D.\ J.\ Thouless, J.\ Phys. C%
{\bf 6} 1181 (1973).
\end{thebibliography}
\end{document}